\documentclass[aps,prb,preprint,groupedaddress]{revtex4-1}

\usepackage{graphicx}
\usepackage{latexsym}
\usepackage{amsmath,amssymb}
\begin{document}

\title{
N\'eel-VBS phase boundary of the extended
$J_1$-$J_2$ model with  
biquadratic interaction
}


\author{Yoshihiro Nishiyama}  
\affiliation{Department of Physics, Faculty of Science,
Okayama University, Okayama 700-8530, Japan}


\date{\today}

\begin{abstract}
The $J_1$-$J_2$ model with the biquadratic (plaquette-four-spin)
interaction
was simulated with the numerical-diagonalization method.
Some limiting cases of this model have been investigated
thoroughly.
Taking the advantage of the extended parameter space,
we survey 
the phase boundary 
separating 
the N\'eel and valence-bond-solid 
phases.
According to the deconfined-criticality scenario,
the singularity of this phase boundary
is continuous, accompanied with
unconventional critical indices.
Diagonalizing the finite-size cluster
with $N \le 36$ spins,
we observe a signature of continuous phase transition.
Our tentative estimate for 
the correlation-length critical exponent is $\nu=1.1(3)$.
In order to elucidate a non-local character of criticality,
we evaluated the Roomany-Wyld $\beta$ function around the critical point.
\end{abstract}


\maketitle

\section{Introduction}

The deconfined criticality\cite{Senthil04,Senthil04b,Levin04}
is arousing much attention recently.\cite{Singh10}
Naively,\cite{Senthil04b}
 the 
(ground-state)
phase transition separating the
N\'eel and 
valence-bond-solid (VBS) phases
is discontinuous, because the adjacent phases possess distinctive
order parameters such as the sublattice magnetization and the
dimer-coverage pattern,
respectively.
However, according to the deconfined-criticality scenario,
the transition is continuous, accompanied with novel critical indices;%
\cite{Motruich04,Tanaka06,Liu05,Dellenschneider06,Ghaemi06,Thomas07,Kim97}
afterward, we make an overview on recent computer-simulation studies.

\begin{figure}
\includegraphics{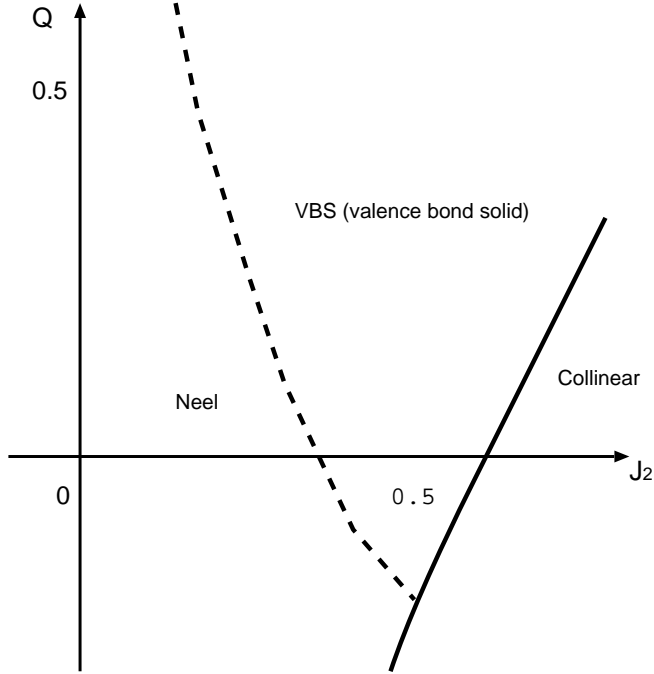}%
\caption{\label{figure1}
A schematic (ground-state) phase diagram for the
$J_1$-$J_2$-$Q$ model (\ref{Hamiltonian})
is presented.
The solid (dashed) line stands for the phase boundary
of discontinuous (continuous) character.
We investigate the phase boundary separating the N\'eel and VBS phases.
Details of the multi-critical point and the VBS-collinear phase boundary 
are uncertain.
}
\end{figure}

In this paper, we investigate
an extended version of the $J_1$-$J_2$ model\cite{Gelfand89,Dagotto89,Schulz92}  
[see Eq. (\ref{Hamiltonian})]
by means of the numerical diagonalization method
for the cluster with $N \le 36$ spins.
The extension of the parameter space, $(J_2,Q)$,
permits us to investigate the above-mentioned criticality
from a global viewpoint.
A schematic phase diagram is presented in
Fig. \ref{figure1}; details are explicated afterward.
As mentioned above, the critical branch 
(dashed line)
is our concern.


The Hamiltonian of the $J_1$-$J_2$ model 
with the biquadratic interaction $Q$
is
given by
\begin{equation}
\label{Hamiltonian}
{\cal H}= J_1
\sum_{\langle ij \rangle} {\bf S}_i \cdot {\bf S}_j 
+J_2
\sum_{\langle \langle ij \rangle \rangle } {\bf S}_i \cdot {\bf S}_j 
-Q
\sum_{[ijkl]}
[
({\bf S}_i \cdot {\bf S}_j -1/4)({\bf S}_k \cdot {\bf S}_l -1/4)+ 
({\bf S}_i \cdot {\bf S}_l -1/4)({\bf S}_j \cdot {\bf S}_k -1/4) 
]
                     .
\end{equation}
Here, the quantum spin-$1/2$ operators $\{ {\bf S}_i \}$ are
placed at each square-lattice point $i$.
The summations,
$\sum_{\langle ij \rangle}$,
$\sum_{\langle \langle ij \rangle \rangle}$,
and 
$\sum_{[ijkl]}$,
run 
over all possible
nearest-neighbor pairs $\langle ij \rangle$,
next-nearest-neighbor pairs
$\langle \langle ij \rangle \rangle $,
and
plaquette spins $[ijkl]$, respectively;
here, 
the arrangement of indices,
$[ijkl]$,
 around a plaquette, $\Box$, 
is ${}^i_j \Box_k^l $.
The parameters, $J_1$, $J_2$, and $Q$,
are the corresponding coupling constants.
Hereafter,
we consider $J_1$ as the unit of energy;
namely, we set $J_1=1$.
Both frustration and biquadratic 
interactions, $J_2$ and $Q$, stabilize the VBS phase.

In some limiting cases (subspaces),
namely, either $Q=0$ or $J_2=0$,
the Hamiltonian (\ref{Hamiltonian}) has been investigated 
extensively so far.
The case of $Q=0$, namely,
the $J_1$-$J_2$ model,\cite{Oitmaa96} has been analyzed
with the series-expansion method\cite{Oitmaa96,Sirker06,Darradi08,Rochter10}
and the numerical diagonalization 
method.\cite{Rochter10,Poilblanc06,Capriotti03,Sindzingre03,Richter10}
(The quantum Monte Carlo method is inapplicable to such a frustrated
magnetism because of the negative-sign problem.)
The N\'eel, VBS, and collinear phases 
appear\cite{Oitmaa96} successively
in the regimes,
$J_2/J_1 \lesssim 0.4$, 
$0.4 \lesssim J_2 / J_1 \lesssim 0.6 $,
and 
$0.6 \lesssim J_2/J_1 $, respectively.
The singularity\cite{Yu11} at $J_2/J_1 \approx 0.4$ ($0.6$)
is continuous (discontinuous\cite{Sirker06}).
On the one hand,
the case of $J_2=0$, namely,
the $J$-$Q$ model\cite{Sandvik07,Melko08} 
($J_1$-$Q$ model in our notation),
admits the use of the quantum Monte Carlo method.
The N\'eel phase turns into the VBS phase 
at a considerably large biquadratic interaction 
$Q/J \approx 25$.\cite{Sandvik07}     
This fact indicates that the antiferromagnetic order
at $J_2=0$ is considerably robust against $Q$.
Nevertheless,
the large-scale quantum Monte Carlo simulation
admits detailed analyses of criticality.
The correlation-length critical exponent is estimated as
$\nu=0.78(3)$
(Ref. \onlinecite{Sandvik07})
and 
$\nu=0.68(4)$
(Ref. \onlinecite{Melko08}).
According to Ref. \onlinecite{Sandvik10},
logarithmic scaling corrections prevent us from taking the thermodynamic limit
reliably.
Possibly because of this difficulty,
it is still controversial whether the singularity
is continuous
or belongs to a discontinuous 
one accompanied with an appreciable latent 
heat\cite{Kotov09,Isaev10,Kotov10,Isaev10b,Kuklov04,Kuklov08,Jian08,Kruger06};
see aforementioned Ref. \onlinecite{Sirker06} as well.
Last, it has be mentioned that
there have been reported a number of intriguing extentions
such as the 
third-neighbor interaction,\cite{Reuther11}
internal symmetry enlargement,\cite{Kawashima07}
and
spatial anisotropy.\cite{Harada07,Nishiyama09,Nishiyama11}
The dimer 
model\cite{Charrier10,Papanikolaou10} 
is also a clue to the study of deconfined criticality.

In this paper,
taking the advantage of the extended parameter space
$(J_2,Q)$ (see Fig. \ref{figure1}), we survey the critical branch
in details.
Scanning the parameter space, we found that around 
an intermediate regime $Q=0.2$,
the finite-size-scaling behavior improves significantly.
Around this optimal regime,
we carry out finite-size-scaling analyses

The rest of this paper is organized as follows.
In Sec. \ref{section2},
we present the simulation results.
Technical details are explicated.
In Sec. \ref{section3},
we address the summary and discussions.

\section{\label{section2}
Numerical results}

In this section, we present the numerical results
for the $J_1$-$J_2$-$Q$ model, Eq. (\ref{Hamiltonian}).
We employ the exact-diagonalization method
for the cluster with $N(=L^2) \le 6^2$ spins.
Our aim is to clarify the nature of the phase boundary
separating the N\'eel and VBS phases.
For that purpose,
we scrutinize the excitation gap
\begin{equation}
\label{excitation_gap}
\Delta E = 
E_{[(0,\pi),0,+,+,+]}-
E_{[(0,0),0,+,+,+]}
              .
\end{equation}
Here, the variable $E_\alpha$ denotes
the ground-state energy within the sector (subspace)
specified by the set of quantum numbers,
$\alpha=(\vec{k},S^z,\pm,\pm,\pm)$.
These quantum numbers
$(\vec{k},S^z,\pm,\pm,\pm)$
are associated with 
the symmetry groups such as
translation along the rectangular edges,
internal-spin $z$-axis rotation
(total-spin conservation along the $z$ axis),
internal-spin inversion
($S^z_i \leftrightarrow -S^z_i$), $\theta=\pi$ lattice rotation,
and lattice inversion, respectively.
These symmetry groups commute mutually,
because we restrict ourselves within the sector with $k_x=0$ and $S^z=0$.
The excitation gap, Eq. (\ref{excitation_gap}),
opens (closes) for the N\'eel (VBS) phase in the thermodynamic limit.
(The degeneracy in the VBS phase corresponds to the 
translationally-invariant
dimer-coverage 
patterns.)
The excitation branch (\ref{excitation_gap}) does not correspond to the first energy gap:
There appear numerous low-lying excitations
due to the spontaneous-symmetry breakings in both N\'eel and VBS phases.
Here, aiming to discriminate these phases clearly,
we look into an excitation branch 
(\ref{excitation_gap}),
which is a key ingredient of the finite-size-scaling analysis.

　　　

\subsection{Survey of the phase boundary 
between the VBS and N\'eel phases}

\begin{figure}
\includegraphics{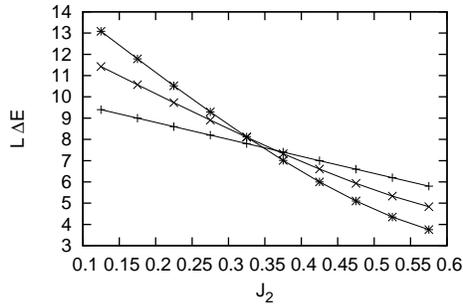}%
\caption{\label{figure2}
The scaled energy gap $L \Delta E$ is plotted for the 
next-nearest-neighbor interaction $J_2$ and the system size
$L=$($+$) 2, ($\times$) 4, and ($*$) 6.  
The biquadratic interaction is fixed to $Q=0.2$.
The data indicate an onset of continuous phase transition around
$J_2 \approx 0.3$.
}
\end{figure}

In Fig. \ref{figure2},
we plot the scaled energy gap $L \Delta E$
for $Q=0.2$, various $J_2$, and  $N=4$, $16$, and $36$.
Here, the variable $L(=\sqrt{N})$ denotes the linear dimension of the cluster.
The intersection point of the curves indicates a location of the
critical point. 
We see that a transition occurs at $J_2 \approx 0.3$.

\begin{figure}
\includegraphics{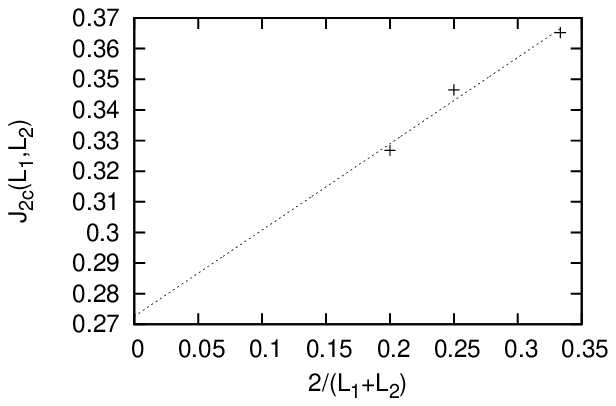}%
\caption{\label{figure3}
The approximate critical point 
$J_{2c}(L_1,L_2)$ (\ref{critical_point})
is plotted for $2/(L_1+L_2)$ with $2 \le L_1 < L_2 \le 6$.
The parameters are the same as those of Fig. \ref{figure2}.
The least-squares fit to these data yields 
$J_{2c}=0.273(12)$ in the thermodynamic limit.
A possible systematic error is considered in the text.
}
\end{figure}

In order to estimate the transition point precisely,
in Fig. \ref{figure3},
we plot the approximate critical point 
$J_{2c}(L_1,L_2)$
for $2/(L_1+L_2)$
with $2 \le L_1 < L_2 \le 6$; the parameters are the same as those of
Fig. \ref{figure2}.
Here, the approximate critical point 
$J_{2c}(L_1,L_2)$
denotes a scale-invariant point
with respect to a pair of system sizes $(L_1,L_2)$.
Namely, it satisfies the relation
\begin{equation}
\label{critical_point}
L_1 \Delta E(L_1)|_{J_2=J_{2c}(L_1,L_2)}=
L_2 \Delta E(L_2)|_{J_2=J_{2c}(L_1,L_2)}
  .
\end{equation}
The least-squares fit to the data of Fig. \ref{figure3}
yields an estimate $J_{2c}=0.273(12)$ in the thermodynamic limit
$L \to \infty$.
Replacing the abscissa scale with $1/L^2$,
we obtain an alternative estimate
$J_{2c}=0.3095 (91)$.
The discrepancy between these estimates,
$\approx 3.7 \cdot 10^{-2}$,
appears to dominate the insystematic (statistical) error, $\approx 10^{-2}$.
Considering the former 
as an indicator of the error margin,
we estimate the transition point $J_{2c}=0.273(37)$.

We carried out similar analyses for various values of $Q$.
The results are shown in 
Fig.
\ref{figure4}.
The error margin is suppressed 
around an intermediate regime $Q \sim 0.1$.
As a matter of fact,
in Fig. \ref{figure5}, 
we show that the scaling behavior becomes optimal
around $Q \approx 0.2$.
(Such a redundancy is a benefit of the parameter-space
extension.)
On the contrary, the error margins get enhanced
around the multi-critical point as well as in
the large-$Q$ regime.

As mentioned in the Introduction,
the limiting cases, $Q=0$ and $J_2=0$, have been studied 
extensively.\cite{Oitmaa96,Sandvik07}
In Ref. \onlinecite{Oitmaa96},
The transition point $(J_2,Q) \approx(0.4,0)$ was 
reported.
This result agrees with ours.
On the contrary,
the transition point\cite{Sandvik07}
 $(J_2,Q) \approx (0,25)$ is not reproduced by the present approach.
This discrepancy may be attributed to the 
logarithmic corrections\cite{Sandvik10}
inherent in the subspace $J_2=0$;
these corrections prevent
us from taking the thermodynamic limit reliably.
Around $Q \approx 0$,
there arise pronounced scaling corrections, which cannot be
appreciated properly;
this regime would not be accessible by
the numerical diagonalization method.
As a matter of fact,
the $Q$-driven VBS state is so unstable
that a considerably weak
antiferromagnetic interaction
$J_1=0.04$ destroys VBS immediately.
This difficulty is remedied by the
inclusion of 
the magnetic frustration $J_2$, which appears to
stabilize VBS significantly (Fig. \ref{figure4}),
and recover an applicability of the diagonalization method.

Last, we make a comment.
The present simulation
cannot rule out a possibility of the discontinuous phase transition.
Even in such a case, the phase boundary determined through
the criterion,
Eq. (\ref{critical_point}),
makes sense,
albeit the value of $L \Delta E$ at the transition point
no longer converges to a universal value (critical amplitude).

\subsection{Correlation-length critical exponent}

In this section, we estimate the correlation-length critical exponent $\nu$.
In Fig. \ref{figure5},
we plot the approximate critical exponent
\begin{equation}
\label{critical_exponent}
\nu(L_1,L_2)=
\frac{\ln (L_1/L_2)}{
  \ln \{\partial_J [L_1\Delta E(L_1)]/\partial_J[L_2\Delta E(L_2)] \}
  |_{J_2=J_{2c}(L_1,L_2)}
                    }
       ,
\end{equation}
for $2/(L_1+L_2)$ and $2 \le L_1 < L_2 \le 6$.
The parameters are the same as those of Fig. \ref{figure2}.
The least-squares fit to these data yields 
$\nu=1.094 (26) $ in the thermodynamic limit.
In order to appreciate possible systematic errors,
we made an alternative extrapolation with the $1/L^2$-abscissa scale.
Thereby, we arrive at $\nu= 1.087(14)$.
In this (optimal) parameter space, $Q=0.2$, the systematic error 
(deviation between different abscissa scales) appears to be negligible.
On the one hand,
at $Q=0.3$, 
we estimated
$\nu=0.901(21)$
and 
$0.988(17)$
through the $1/L$- and $1/L^2$-extrapolation schemes, respectively.
The discrepancy (systematic error) appears to be enhanced
by the scaling (possibly, logarithmic\cite{Sandvik10}) corrections.
At $Q=0.1$,
via the $1/L$- and $1/L^2$-extrapolation schemes, 
we arrive at
$\nu=1.316(87)$
and 
$1.198(55)$, respectively.
Both systematic and insystematic errors get enhanced
in the proximity to the multicritical point.
Considering these deviations of various characters
as an indicator of the error margin,
we 
estimate the critical exponent as
\begin{equation}
\nu=1.1(3)  .
\end{equation}
A validity of the scaling analysis is 
considered in the next section.

\subsection{Non-local behavior of the $\beta$ function}

In Fig. \ref{figure6},
we plot the $\beta$ function,
$\beta^{RW}_{L_1 L_2}(J_2)$,
 for $Q=0.2$,
various $J_2$, and 
        $(L_1,L_2)=$ $(+)$ $(2,4)$,
           $(\times)$ $(4,6)$,
and $(*)$ $(2,6)$.
Here, 
we calculated the $\beta$ function with the Roomany-Wyld
formula\cite{Roomany80}
\begin{equation}
\label{beta_function}
\beta^{RW}_{L_1 L_2}(J_2)
=
\frac{1+ \ln(\Delta E(L_1)/\Delta E(L_2))/\ln(L_1/L_2)}
{
\sqrt{ \partial_{J_2} \Delta E(L_1)
       \partial_{J_2} \Delta E(L_2)/
     \Delta E(L_1)/\Delta E(L_2) }
}
, 
\end{equation}
for a pair of system sizes $(L_1,L_2)$.

The $\beta$ function
elucidates a non-local character of the criticality.
In the vicinity of
the critical point $J_{2c}$,
the $\beta$ function 
falls into a linearized formula
\begin{equation}
\beta(J_2)= \frac{1}{\nu} (J_{2c}-J_2)
                ,
\end{equation}
with the correlation-length critical exponent $\nu$.
In Fig. \ref{figure6}, we present the theoretical prediction
$\beta = (0.273-J)/1.1$ 
(dotted line);
here, the critical parameters, $J_{2c}=0.273$ and $\nu=1.1$,
are estimated in the preceding sections.
The behavior of $\beta^{RW}_{L_1 L_2}$ is consistent with
the theoretical prediction for a considerably wide range of $J_2$,
validating out treatment.
In particular, the linearity of $\beta^{RW}_{L_1 L_2}$
is remarkable, suggesting 
that scaling corrections are eliminated satisfactorily,
at least,  around $Q=0.2$.
Possibly,
the $J_2$-driven phase transition,
rather than the $Q$-driven transition, is reasonable
in the sense of the renormalization-group flow.

\subsection{Ground-state energy: An indication of the VBS-collinear phase transition}

In Fig. \ref{figure7},
we present the ground-state energy per unit cell, $E_g /N$,
with $N=36$ for $Q=0.2$ and various $J_2$.
From the figure, we observe a cusp-like anomaly around $J_2 \approx 0.7$.
This anomaly indicates an onset of the first-order phase transition 
between the VBS and collinear phases.
Making similar analyses for various values of $Q$, 
we obtained a schematic feature of the
VBS-collinear phase boundary 
as shown in Fig. \ref{figure1}.
Character of the multicritical point is unclear.


\section{
\label{section3}
Summary and discussions}

\begin{figure}
\includegraphics{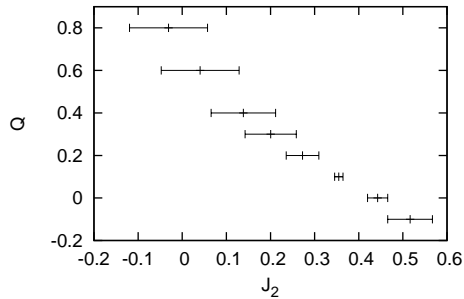}%
\caption{\label{figure4}
The phase boundary between the N\'eel and VBS phases (critical branch)
is presented.
}
\end{figure}

\begin{figure}
\includegraphics{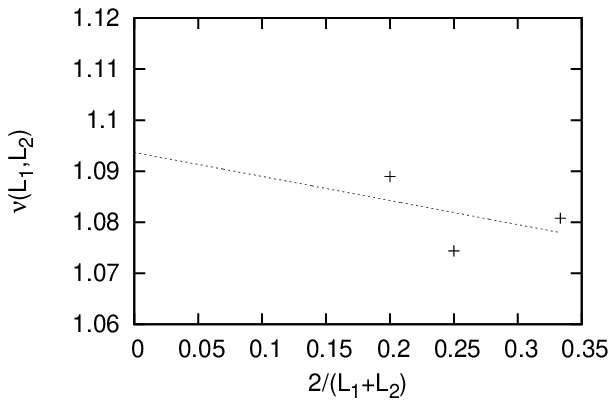}%
\caption{\label{figure5}
The approximate critical exponent
$\nu(L_1,L_2)$ (\ref{critical_exponent})
is plotted for $2/(L_1+L_2)$ with $2 \le L_1 < L_2 \le 6$.
The parameters are the same as those of Fig. \ref{figure2}.
The least-squares fit to these data yields $\nu=1.094(26)$ in the thermodynamic limit.
A possible systematic error is considered in the text.
}
\end{figure}

\begin{figure}
\includegraphics{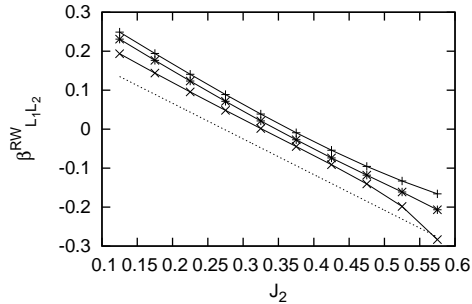}%
\caption{\label{figure6}
The beta function
$\beta^{RW}_{L_1 L_2}$ (\ref{beta_function})
is plotted for $Q=0.2$,
 various $J_2$,
and  $(L_1,L_2)=$ 
$(+)$ $(2,4)$,
$(\times$) $(4,6)$, and 
$(*)$ $(2,6)$. 
For a comparison,
we present a theoretical prediction $\beta=(0.273-J_2)/1.1$ (dotted line); see text for details.
We see that the $\beta$ function follows the theoretical prediction
for a wide range of $J_2$.
}
\end{figure}

\begin{figure}
\includegraphics{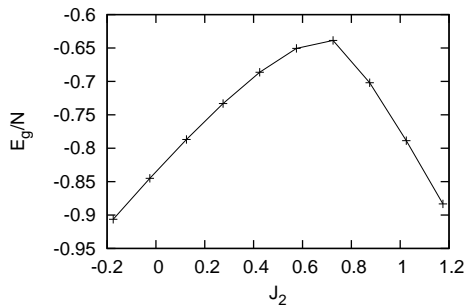}%
\caption{\label{figure7}
The ground-state energy per unit cell $E_g/N$ with $N=36$ is presented
for $Q=0.2$ and various $J_2$.
}
\end{figure}

The phase diagram of
the extended $J_1$-$J_2$ model
with the biquadratic interaction $Q$,
Eq. (\ref{Hamiltonian}),
was investigated
by means of
the numerical diagonalization method for $N \le 36$.
Taking the advantage of the extended parameter space, $(J_2,Q)$,
we surveyed the 
critical branch separating
the
N\'eel and VBS phases (Fig. \ref{figure4}) in particular.
The finite-size corrections appear to be
suppressed
around
an intermediate regime $Q \approx 0.2$.
Encouraged by this finding,
we analyzed the criticality around $Q \approx 0.2$
with the finite-size-scaling method.
As a result, we estimate the correlation-length
critical exponent $\nu=1.1(3)$.
A non-local feature of the $\beta$ function is accordant with
the
theoretical prediction
(Fig. \ref{figure6}),
giving support to
our treatment.

As a reference,
we overview related studies.
For the $J$-$Q$ model
(square-lattice antiferromagnet with the biquadratic interaction),
the
estimates, $\nu=0.78(3)$ (Ref. \onlinecite{Sandvik07}) and 
$\nu=0.68(4)$ (Ref. \onlinecite{Melko08}), were reported.
As for the spatially-anisotropic spin systems, 
the exponent $\nu=0.80(15)$ (Ref. \onlinecite{Nishiyama09}) 
and 
$\nu=0.92(10)$ (Ref. \onlinecite{Nishiyama11})
were
obtained.
These results are to be compared with 
$\nu=0.7112(5)$ (Ref. \onlinecite{Campostrini02})
for 
the $d=3$ Heisenberg universality class.
Our result indicates a tendency toward an enhancement for the correlation-length critical exponent,
as compared to that for the $d=3$ Heisenberg universality class.

According to Ref. \onlinecite{Xu09},
information on the deconfined criticality can be drawn from 
the details of the multi-critical point.
This viewpoint would provide an evidence
whether the criticality belongs to
the deconfined criticality;
at present, it is not conclusive whether the N\'eel-VBS
transition belongs to the deconfined criticality. 
This problem will be addressed in future study.

\end{document}